\documentclass[draf]{aa} 

\usepackage{graphicx}
\usepackage{txfonts}
\usepackage{xcolor}
\usepackage{ulem}
\usepackage[breaklinks=true]{hyperref} 

\idline{A\&A, 706, A182 (2026)}
\doi{https://doi.org/10.1051/0004-6361/202557509}

\begin{document}

   \title{A single frequency approach to nonequilibrium modeling of the chromosphere}

   \author{W. Ruan
          \and
          D. Przybylski
          \and
          R. Cameron
          \and
          S. K. Solanki
          }

   \institute{Max-Planck Institute for Solar System Research, 37077 Göttingen, Germany\\
              \email{ruanw@mps.mpg.de}
             }

  \abstract
   {The solar chromosphere is a region where radiation plays a critical role in energy transfer and interacts strongly with the plasma. In this layer, strong spectral lines, such as the Lyman lines, contribute significantly to radiative energy exchange. Due to the long ionization/relaxation timescale, departures from local thermodynamic equilibrium (LTE) become significant in the chromosphere. Accurately modeling this layer therefore requires one to solve the non-LTE radiative transfer for the Lyman transitions.}
   {We present an updated version of the MURaM code to enable more accurate simulations of chromospheric hydrogen level populations and temperature evolution.}
   {In the previous extension, a non-LTE equation of state, collisional transitions of hydrogen, and radiative transitions of non-Lyman lines were already implemented in the code. Building on this, we have now incorporated radiative transfer for the Lyman lines to compute radiative rate coefficients and the associated radiative losses. These were used to solve the population and temperature evolution equations, rendering the system self-consistent. To reduce computational cost, a single-frequency approximation was applied to each line in the numerical solution of the radiative transfer problem.}
   {The extended model shows good agreement with reference solutions from the Lightweaver framework, accurately capturing the radiative processes associated with Lyman lines in the chromosphere. The extension brings the simulated hydrogen level populations in the deep chromosphere closer to detailed radiative balance, while those in the upper chromosphere remain significantly out of balance, consistent with the expected conditions in the real solar atmosphere. Convergence tests show that the module can accurately capture the evolution of temperature and hydrogen level populations with simulation time steps constrained by the Courant–Friedrichs–Lewy (CFL) condition.}
   {The extension enables the MURaM code to accurately capture chromospheric dynamics. Its robust performance under large simulation time steps renders it particularly well suited for high-resolution, three-dimensional simulations.}


   \maketitle
   
   \nolinenumbers

\section{Introduction}

The chromosphere is the region located above the solar photosphere, characterized by strong brightness in the H$\alpha$ waveband seen in emission during eclipses. 
In one-dimensional (1D) empirical models of the solar atmosphere, the temperature in the chromosphere gradually increases with altitude, from approximately 4000 K to 8000 K at around 2000 km above the bottom of the photosphere \citep{Avrett2008ApJS}.
High-resolution observations and three-dimensional (3D) radiation-magnetohydrodynamics (MHD) simulations show a considerably more complex picture. The chromosphere is structured and dynamic, and the transition region highly corrugated.
In the chromosphere, the energy transmitted from the solar interior is redistributed, with the majority radiated into space, and a smaller portion transported to the corona \citep{Jess2009Sci}. Understanding the chromosphere, the distribution of energy within it, and its coupling to the corona is essential for addressing key questions in solar physics, such as coronal heating and the driving of the solar wind \citep{Cranmer2019ARA&A,VanDoorsselaere2020SSRv}. Additionally, the chromosphere itself has many interesting issues to be solved, such as its heating and the formation of fine structures such as spicules. A recent review of the chromosphere has been provided by \citet{Carlsson2019ARA&A}, offering a comprehensive overview of researchers' current understanding and outstanding questions in the field.

Simulating the evolution of the lower solar atmosphere is challenging, partly because radiation plays a critical role in energy transfer and interacts strongly with the plasma. Thus the simulation codes must solve the radiative transfer (RT) equation to model the radiative energy exchange processes \citep{Leenaarts2020LRSP}.
In the photosphere, which is close to local thermodynamic equilibrium (LTE), the level populations follow Saha-Boltzmann statistics. The source function of the main spectral band, the continuum spectrum, can then be approximated by the temperature-dependent Planck function.
The chromosphere is particularly challenging due to non-local thermodynamic equilibrium (NLTE) effects.
In the chromosphere, strong spectral lines such as the \ion{Ca}{II}, \ion{Mg}{II}, and H\,\textsc{i} Lyman-$\alpha$ lines play a key role in radiative energy transfer \citep{Vernazza1981ApJS}.
Due to long ionization and recombination timescales in the chromosphere, the level populations of the atoms and ions, required for the solution of the RT equation, become decoupled from the local temperature \citep{Carlsson2002ApJ}. Since the evolution of these populations depends in turn on the radiation field, this interdependent relationship makes the modeling of radiative energy exchange complicated and potentially global.

In terms of computational effort, fully solving the RT equation to obtain an accurate, frequency-dependent radiation field for spectral lines is highly expensive -- not only due to the large number of frequency points required, but also because of the effects of partially coherent scattering  \citep{Leenaarts2020LRSP}.
Due to the high computational cost, fully solving 3D RT is currently only feasible in spectral synthesis, such as with MULTI3D \citep{Leenaarts2009ASPC} and PORTA \citep{Stepan2013A&A}.
Time-dependent calculations including a detailed treatment of the radiation transfer are typically restricted to 1D simulations (e.g., \citealp{Carlsson1997ApJ,Kasparova2009A&A}).

In chromospheric simulations, the primary emphasis is typically placed on the evolution of the plasma state.
In particular, the evolution of hydrogen receives significant attention due to its high abundance and dominant role in plasma dynamics. When modeling the interaction between matter and radiation, using simplified radiation fields is an effective approach to reduce computational complexity.
One widely used method is that of \citet{Sollum1999}, which approximates the radiation field of hydrogen non-Lyman lines using a Planck function evaluated at a specified local radiation temperature. This method has been implemented in several simulation codes, including CO5BOLD \citep{Leenaarts2006ASPC}, the Oslo Stagger Code \citep{Leenaarts2007A&A}, Bifrost \citep{Golding2010, Gudiksen2011A&A}, and MURaM \citep{Przybylski2022A&A}, which incorporate NLTE ionization effects in modeling chromospheric plasma.
An alternative approach of simplifying the radiation field, utilizing an escape probability method, was suggested by \citet{Judge2017ApJ}.

The Lyman series, particularly the Lyman-$\alpha$ line, plays a significant role in cooling the chromosphere and thus cannot be ignored in NLTE simulations. 
One approach to modeling the Lyman series is to assume detailed radiative balance, in which the local upward radiative rate equals the downward radiative rate. Under the assumption of radiative balance, population changes due to radiative excitation and de-excitation are neglected (e.g., \citealp{Leenaarts2007A&A,Przybylski2022A&A}).
Detailed radiative balance is a good approximation for Lyman-$\alpha$ in the lower chromosphere, but it does not apply in the upper chromosphere \citep{Vernazza1981ApJS, Carlsson2012A&A}. 
In the upper chromosphere, radiative de-excitation generally dominates over radiative excitation, with Lyman-$\alpha$ photons carrying a substantial amount of energy into space. Additionally, photons emitted from the transition region are reabsorbed in the upper chromosphere, potentially influencing its thermal and dynamical evolution.
\citet{Golding2016ApJ} introduced RT of the Lyman-$\alpha$ transition into the Bifrost code and demonstrated that the resulting hydrogen ionization degree near wavefronts differs significantly from that obtained under the assumption of radiative detailed balance. To accurately capture the hydrogen population dynamics in the upper chromosphere, computing the radiation field of the Lyman series by solving the RT equation might be necessary.

In this paper, we present an extension of the MURaM code. Building on the chromospheric extension introduced in \citet{Przybylski2022A&A}, we develop a patch to incorporate the RT treatment, initially restricted to modeling the Lyman series.
The treatment can easily be expanded to model other species that are important for the chromospheric energy balance, such as He, \ion{Ca}{II}, and \ion{Mg}{II}.
This involves expanding the treatment of the RT to calculate the radiation field utilizing the 3D short-characteristics solver.

We aim to obtain simulated hydrogen population distributions that more closely reflect conditions in the real solar atmosphere, where the heating or cooling due to the radiation field is consistent with that used to treat the atomic populations.
In the upper chromosphere, the inclusion of RT enables the simulation of important processes, whereby photons emitted from the transition region and from shock fronts are absorbed by the upper chromosphere.
In the lower chromosphere, using radiative rate coefficients from RT can help align the population with the expected detailed radiative balance. 
In simulations assuming detailed radiative balance, excitation and de-excitation processes are dominated by collisions, with radiative rate coefficients set to zero, thereby imposing no constraints on population evolution. As a result, assuming detailed radiative balance may lead to population simulation results that fall outside of the detailed radiative balance.
To reduce computational cost, a single-frequency approach was employed for treating the transitions, following \citet{Golding2016ApJ}, in which the detailed spectral line profile is neglected, and both opacity and emissivity are assumed to be uniform across the specified passband. 

The theoretical model of the single-frequency approach, along with the numerical method used to solve the RT, is described in Sect.~\ref{sec:method}.
In Sect.~\ref{sec:result}, the accuracy of the treatment in computing radiative rate coefficients and radiative heating or cooling, its impact on the low atmosphere in simulation, the impact of the solver and the Doppler effect, as well as the influence of time step selection on the simulation results are investigated. Finally, a summary is given in Sect.~\ref{sec:summary}.

\section{Method} \label{sec:method}

This section presents the governing equations for NLTE RT in the MURaM code, along with the numerical methods employed to solve them.
For a detailed description of other aspects of the MURaM code, please refer to \citet{Vogler2005A&A}, \citet{Rempel2017ApJ} and \citet{Przybylski2022A&A}.

\subsection{Time-dependent treatment of hydrogen ionization} \label{sec:pops-equations}

The nonequilibrium (NE) treatment of hydrogen ionization for the chromosphere is incorporated from the version of the MURaM code presented in \citet{Przybylski2022A&A}.
The NE treatment considers five energy levels of the hydrogen atom, as well as protons and H$_2$. Additionally, H$_2^+$ and H$^-$ are treated in chemical equilibrium, and non-hydrogen species are treated in LTE.
The hydrogen population updates at each time step account for the effects of advection, bound-bound transitions, bound-free transitions, and chemical reactions, in which the contribution from advection, bound-bound transitions and bound-free transitions is expressed as
\begin{equation}
    \frac{\partial n_{i}}{\partial t} + \nabla \cdot (n_i \mathbf{v}) = \sum_{j \neq i} n_{j} (C_{ji} + R_{ji}) - n_{i} \sum_{j \neq i} (C_{ij}+R_{ij}) , \label{eqn:Rate}
\end{equation}
where $n$ represents number density, $i$ and $j$ denote energy levels, $C$ is the collisional rate coefficient, and $R$ is the radiative rate coefficient. 
The problem is split, with the population advection calculated using the partial donor cell method presented in \citet{Zhang2019ApJS}.

The temperature- and electron-dependent collisional rates were calculated using approximate cross-sections for hydrogen-electron collisions from \citet{Johnson1972ApJ}, with the formula referenced in \citet{Leenaarts2007A&A}.
The radiative rates for the Balmer, Paschen, and Brackett series were calculated following the prescription presented in \citet{Sollum1999}, where the radiation field is expressed as the Planck function of a defined radiation temperature. 
The Lyman series were treated with a different method, which is introduced later.

The radiative cooling term is given by
\begin{equation}
    Q_{\mathrm{rad}} = Q_{\mathrm{phot}} + Q_{\mathrm{lines}} + Q_{\mathrm{thin}} + Q_{\mathrm{back}}, \label{eqn:Qrad}
\end{equation}
where $Q_{\mathrm{phot}}$ represents the contribution from photospheric radiation, $Q_{\mathrm{lines}}$ accounts for the influence of strong chromospheric spectral lines, $Q_{\mathrm{thin}}$ corresponds to coronal optically thin losses and is applicable in regions where temperatures exceed 10,000 K and pressures fall below a user-defined threshold, and $Q_{\mathrm{back}}$ represents the back-heating of chromospheric plasma by coronal radiation.
The photospheric radiation was derived using multigroup RT, accounting for the effects of scattering \citep{Ludwig1992Thesis}.
The radiative heating or cooling contribution from spectral lines $Q_{\mathrm{lines}}=Q_{\mathrm{H}} + Q_{\mathrm{Mg}} + Q_{\mathrm{Ca}}$ includes the contributions from strong spectral lines of hydrogen, magnesium, and calcium. The terms $Q_{\mathrm{Mg}}$ and $Q_{\mathrm{Ca}}$ were computed following the recipe of \citet{Carlsson2012A&A}, in which only radiative losses are considered. The cooling rate in this recipe is expressed as a function of a defined column mass, electron number density, temperature, and element number density. The calculation method for $Q_{\mathrm{H}}$ is presented later.
The coronal optically thin losses, $Q_{\mathrm{thin}} = - n_e n_{H} \Lambda(T)$, is a function of the electron number density, $n_e$, hydrogen number density, $n_H$, and a temperature-dependent curve, $\Lambda(T)$, obtained from the CHIANTI atomic database \citep{Landi2012ApJ}.
The back-heating by coronal radiation, $Q_{\mathrm{back}}$, was calculated using RT, following the approach presented in \citet{Carlsson2012A&A}. In this approach, the emissivity, $Q_{\mathrm{thin}}/4\pi$, and the opacity at the ionization edge of helium are incorporated into the RT calculations.

In the previous version of the code presented in \citet{Przybylski2022A&A}, the Lyman series is treated assuming detailed balance in radiative transitions, with $R_{ij} = R_{ji} = 0$ applied when updating populations. The heating or cooling rate from hydrogen lines ($Q_{\mathrm{H}}$) is calculated using the recipe of \citet{Carlsson2012A&A}, i.e., the same approach as is used for magnesium ($Q_{\mathrm{Mg}}$) and calcium ($Q_{\mathrm{Ca}}$).
The radiative detailed balance assumption deviates significantly from real conditions in the upper chromosphere. Therefore, in this extension of the code, the assumption is discarded when treating the Lyman series. Instead NLTE RT is explicitly solved for the two strongest lines in Lyman series: Lyman-$\alpha$ and Lyman-$\beta$. The radiative rates obtained from the RT solution are used to update hydrogen populations, and $Q_{\mathrm{H}}$ is the total radiative gain or loss from the two lines.
Other Lyman series lines are much weaker than the two strongest lines, and their impact on chromospheric evolution is relatively minor \citep{Carlsson2012A&A}. To reduce numerical effort, they are still treated under the detailed radiative balance assumption.

\begin{figure*} [ht]
    \centering
    \includegraphics[width=1.\linewidth]{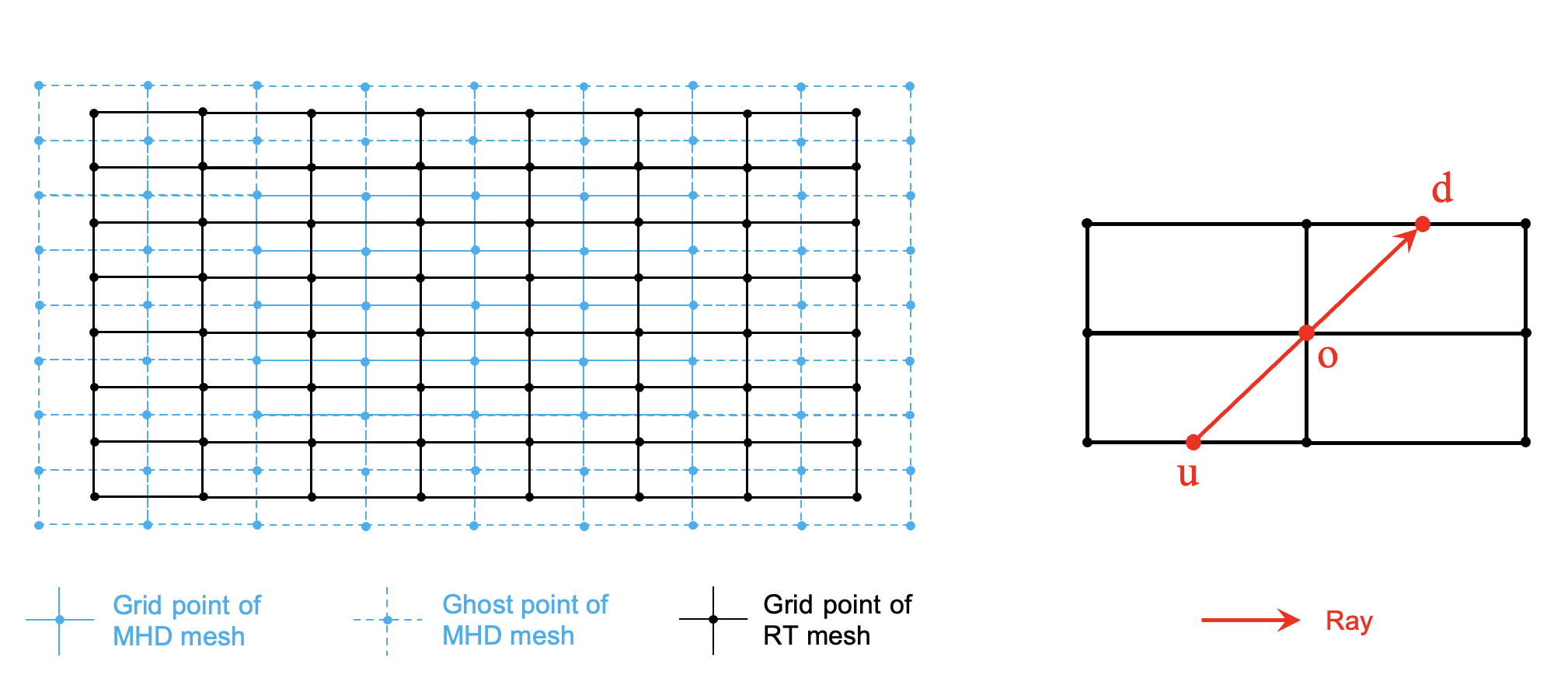}
    \caption{Schematic illustration of the mesh concerning RT. The left panel shows the computational meshes  employed for solving the governing equations in the MURaM code, while the right panel illustrates the relationship between the radiation ray and the mesh structure.
    The MHD mesh is employed for solving the MHD equations, whereas the RT mesh is used for solving the RT equation. Both meshes are uniform and Cartesian, with the RT grid points positioned at the centers of the cells formed by adjacent MHD grid points. Information exchange between the two meshes is accomplished through bi-linear or tri-linear interpolation.
    Rays are involved in solving the RT equation using the short-characteristics method, in which the ray intensity at the grid points ($\mathrm{o}$) is the quantity being solved for in MURaM. The solution involves the source functions at the upwind ($\mathrm{u}$), downwind ($\mathrm{d}$), and grid points, along with the ray intensity at the upwind point.
    The rays typically do not intersect with the grid vertices. Linear or bi-linear interpolation was employed to compute the source function, opacity, and ray intensity at the upwind and downwind points.
    }
    \label{fig:mesh}
\end{figure*}
\subsection{Theoretical model of NLTE radiative transfer in the Lyman series} \label{sec:model-RT}

Radiative transfer solves the time-independent, frequency- and direction-dependent transport equation
\begin{equation}
    \frac{\mathrm{d} I_{\nu}}{\mathrm{d} \tau_{\nu}} = S_{\nu} - I_{\nu}, \label{eqn:RT}
\end{equation}
where $I_{\nu}$, $\tau_{\nu}$, and $S_{\nu}$ represent the ray intensity, optical depth, and source function in a given direction, respectively.
For NLTE RT in the Lyman series, the source function depends primarily on the hydrogen level populations. These spectral lines span a finite bandwidth, and their profiles exhibit significant frequency-dependent variation. Ideally, RT should be solved across multiple frequencies to accurately capture this behavior. However, due to the high computational cost of such calculations, we adopted a single-frequency approximation for the lines, following the approach of \citet{Golding2016ApJ}.

In the single-frequency approach, spectral line-related photons are assumed to be evenly distributed within a given bandwidth, $W_{\nu}$, centered at $\nu_0$. 
The frequency-averaged optical depth, $\tau_{\bar{\nu}}$, is given by the integral of the frequency-averaged opacity $\chi_{\bar{\nu}}$ along the line of sight:
\begin{eqnarray}
    \tau_{\bar{\nu}} (s) &=& \int_{-\infty}^{s} \chi_{\bar{\nu}} (s')\ \mathrm{d} s', \label{eqn:Tau} \\
    \mathrm{d} \tau_{\bar{\nu}} &=& \chi_{\bar{\nu}}\ \mathrm{d} s , \label{eqn:Dtau}
\end{eqnarray}
where $s$ represents the distance along the line of sight. 
The frequency-averaged opacity and source function are defined, respectively, as follows:
\begin{eqnarray}
\chi_{\bar{\nu}} &=& \frac{h \nu_0}{4 \pi} \frac{n_l B_{lu}}{W_{\nu}}, \label{eqn:Chi} \\
S_{\bar{\nu}} &=& \frac{h \nu_0}{4 \pi} \frac{n_u A_{ul}}{W_{\nu} \chi_{\bar{\nu}}}, \label{eqn:S}
\end{eqnarray}
where $h$ is Planck’s constant, $l$ and $u$ represent the lower and upper levels of the transition, $n_l$ the population density of the lower level, $n_u$ the population density of the upper level, $B_{lu}$ the Einstein coefficient for radiative excitation, and $A_{ul}$ the Einstein coefficient for spontaneous de-excitation.
Stimulated emission is significantly weaker than spontaneous emission for these lines and is therefore neglected in the model.
The direction-dependent, frequency-averaged ray intensity, $I_{\bar{\nu}}$, is obtained by substituting $\chi_{\bar{\nu}}$ and $S_{\bar{\nu}}$ into the transport equation (Eqn.~\ref{eqn:RT}) and solving it.
Once the radiation field is known, the heating or cooling associated with this spectral line can be determined. The heating rate is given by
\begin{equation}
Q = h \nu_0 (n_l R_{lu} - n_u R_{ul}), \label{eqn:Q}
\end{equation}
where the downward radiative rate coefficient is $R_{ul}=A_{ul}$, and the upward radiative rate coefficient depends on the frequency-averaged mean intensity, $J_{\bar{\nu}}$: 
\begin{equation}
R_{lu} = B_{lu} J_{\bar{\nu}} = \frac{B_{lu}}{4 \pi} \oint I_{\bar{\nu}}\ \mathrm{d} \Omega, \label{eqn:Rlu}
\end{equation}
where $\Omega$ is a solid angle. 

We adopted line center frequencies of $\nu_0 = c / (121.5,\mathrm{nm})$ for the Lyman-$\alpha$ line and $\nu_0 = c / (102.6,\mathrm{nm})$ for the Lyman-$\beta$ line. The bandwidths of the transitions are user-defined parameters, with default values of $W_{\nu} = 1.0 \times 10^{12}\,\mathrm{Hz}$ for the Lyman-$\alpha$ line and $W_{\nu} = 1.25 \times 10^{12}\,\mathrm{Hz}$ for the Lyman-$\beta$ line. The single-frequency approach with the default bandwidths gives accurate radiative rate coefficients in the test, as demonstrated in Section \ref{sec:one-frequency}. 
The regions of the solar atmosphere in which the NLTE treatment was applied were also user-defined. This was implemented in order to avoid an overlap with cooling from the multigroup losses in the photosphere. Additionally, the solution must be smoothly joined with the LTE photosphere. In order to achieve this, the
non-LTE RT is solved in regions where $p < P_{\mathrm{NLTE\text{-}RT}}$, and the heating or cooling rates and radiative rate coefficients are applied in regions where $p < P_{\mathrm{RT\text{-}USE}}$.
By default, $P_{\mathrm{NLTE\text{-}RT}} = 1 \times 10^5\,\mathrm{dyn\,cm^{-2}}$ and $P_{\mathrm{RT\text{-}USE}} = 1 \times 10^4\,\mathrm{dyn\,cm^{-2}}$, corresponding to typical pressures in the photosphere \citep{Avrett2008ApJS}.

\subsection{Numerical radiative transfer} \label{sec:numerical-RT}

The implementation of numerical RT for the Lyman lines was guided by the principles of simplicity, computational efficiency, and robustness. For typical MURaM radiative MHD simulations, extreme events including strong shocks and eruptive events up to the scale of small flares can occur. The radiation field and populations must be $100\%$ convergent, and radiative instabilities that can feed back into the energy through heating or cooling must be avoided.

The MURaM code utilizes two sets of meshes: one for numerically solving the MHD equations (MHD mesh) and the other for RT equations (RT mesh). Both meshes are uniform and Cartesian. The grid points of the RT mesh are positioned at the center of the cuboids formed by adjacent grid points of the MHD mesh (left panel of Fig.~ \ref{fig:mesh}). 
The populations updated and stored in the MHD mesh are used to solve NLTE RT, with the goal of obtaining the radiative heating or cooling rate ($Q$) and radiative rate coefficients ($R_{lu}$ and $R_{ul}$) at the MHD mesh.
The steps are as follows:
\begin{itemize}
    \item [1)] Obtain populations ($n_l$ and $n_u$) at the RT mesh from the MHD mesh via bi-linear or tri-linear interpolation.
    \item [2)] Calculate the opacity ($\chi$) and source function ($S$) at the RT mesh.
    \item [3)] Solve the RT equation to obtain $J$ at RT mesh.
    \item [4)] Compute $Q$ at the RT mesh.
    \item [5)] Interpolate $Q$ and $J$ to the MHD mesh.
    \item [6)] Calculate $R_{lu}$ and $R_{ul}$ from $J$, $n_l$, and $n_u$ at the MHD mesh.
\end{itemize}
The RT mesh was introduced to enhance robustness. When Q and J are interpolated from the RT mesh to the MHD mesh, their spatial distributions become smoother, thereby reducing the likelihood of numerical instabilities, which could compromise the stability of the simulation. In addition, the existing multigroup RT for photospheric radiation is solved on the RT mesh. Solving the Lyman RT on the same mesh allows us to reuse many of the existing functions, thereby preventing unnecessary code expansion and keeping the implementation streamlined.
A four-step Runge-Kutta-like method was used for time integration in solving the MHD equations \citep{Jameson2017AIAAJ}. 
The complete form of the NE equation of state, including an update of the rate equation, was solved only for the first step. A simplified solution, which conserves charge, hydrogen nucleus number, and energy was solved for the other steps.

The RT problem was solved using the short characteristics method, employing the type A quadrature scheme described in \citet{Carlson1963} for angular integration.
As was noted earlier, the RT is solved on the RT mesh, where our objective is to determine $J$ at each grid point. For this purpose, we selected 24 ray directions (three angular bins per quadrant) following \citet{Carlson1963}, sequentially computed the ray intensity spatial distribution for each direction, and finally averaged these intensity distributions to obtain $J$. The computation of the intensity distribution of a direction requires an iterative solution: at each iteration, the intensity at every grid point is updated based on the current values at adjacent points, and the process continues until convergence is reached. The short characteristics method was employed to perform these intensity updates. The details of the short characteristics method and the iterative scheme are described below.

In the short characteristics method, the local ray intensity is updated based on the intensities at nearby grid points.
The equation of ray intensity can be expressed as
\begin{equation}
I_{i} (\tau_{o,i}) = I_i (\tau_{u,i}) \mathrm{e}^{-(\tau_{o,i} -\tau_{u,i})} + \int_{\tau_{u,i}}^{\tau_{o,i}} S_i (t) \mathrm{e}^{t-\tau_{o,i}} \ \mathrm{d} t, \label{eqn:I}
\end{equation}
where $i$ represents the direction index, $\tau_{o,i}$ is the optical depth of the $i$-th ray at the grid point, and $\tau_{u,i}$ is the optical depth at the upwind point. The upwind point, $u$, is different for different directions.
Here, the second-order scheme BESSER is used as the default solver to numerically integrate the source function, assuming a monotonic quadratic Bézier variation, given by
\begin{equation}
\int_{\tau_{u}}^{\tau_{o}} S (t) \mathrm{e}^{t-\tau_o} \ \mathrm{d} t = \Delta I = \Psi_u S (\tau_{u}) + \Psi_o S (\tau_{o}) + \Psi_c c_{\mathrm{u}}, \label{eqn:Besser}
\end{equation}
where $c_{\mathrm{u}}$ denotes the source function at a specified control point, and $\Psi_u$, $\Psi_o$ and $\Psi_c$ are interpolation weights. 
As a second-order scheme, the value of $c_{\mathrm{u}}$ depends on $S (\tau_{u})$, $S (\tau_{o})$, and downwind point source function $S (\tau_{d})$, with the methods for computing $c_{\mathrm{u}}$ and the weights are detailed in \citet{Stepan2013A&A}.
The spatial relationship between rays and points is depicted in Fig.~ \ref{fig:mesh}. 
As is shown in the figure, the ray does not coincide with the mesh lines, so the upwind and downwind points of the ray generally do not coincide with the grid points. This requires interpolation to determine the source function and ray intensity at these points.
Alternatively, a first-order linear solver \citep{Kunasz1988JQSRT} is available for source function integration.

The solution of ray intensity was obtained using the Gauss–Seidel iterative method, in which the updated values are used within each iteration, based on the following expression:
\begin{equation}
I^{\mathrm{new}} (\tau_{o}) = \Delta I + I^{\mathrm{old/new}} (\tau_{u}) \mathrm{e}^{-\Delta \tau_{ou}}, \label{eqn:iter}
\end{equation}
where $\Delta I$ is given by Eqn.~\ref{eqn:Besser} and is computed in advance prior to the iteration, and $\Delta \tau_{ou}$ is computed in advance as well.
In each iteration step, we updated $I$ at all grid points in the region of interest according to Eqn.~\ref{eqn:iter}. 
The code employs MPI parallel computing, in which the simulation box is divided into multiple regions, with each processor responsible for managing the data within one region. When computing the ray intensity at grid points located at the boundary of a processor’s domain, the upwind point may fall outside its domain, depending on the ray direction, leaving the processor without the necessary intensity data. Consequently, in each iteration, cross-processor communication is required to obtain the ray intensities at points outside the domain. 
The grid points are swept from the upwind side. If the upwind point falls outside the domain, $I (\tau_{u})$ is assigned the value from the preceding iteration ($I^{\mathrm{old}} (\tau_{u})$); otherwise, the updated value is employed ($I^{\mathrm{new}} (\tau_{u})$).
For further details on data exchange related to ray intensity, refers to \citet{Vogler2003}.
The iteration process continues until the error falls below a predefined threshold at all grid points in the region of interest:
\begin{equation}
\max \left ( \frac{I^{\mathrm{new}} - I^{\mathrm{old}}}{\max(I_{\mathrm{min}}, I^{\mathrm{old}})} \right ) < 5 \times 10^{-5}, \label{eqn:threshold}
\end{equation}
where $I_{\mathrm{min}}$ is a defined small value used to prevent division by zero. 
Radiative transfer is solved in the regions where $\Delta \tau_{ou}\geq\Delta \tau_{\mathrm{min}}$, where $\Delta \tau_{\mathrm{min}}$ is a user-specified parameter that allows control over whether RT is also solved in the optically thin corona.
In regions where $\Delta \tau_{ou}<\Delta \tau_{\mathrm{min}}$, the ray intensity $I$ is set to zero, as radiation absorption can be neglected due to the low opacity.

The mean intensity required in Eqn.~\ref{eqn:Rlu} was computed as the average of the ray intensities over 24 directions:
\begin{equation}
J = \sum_i \omega_i I_i, \label{eqn:J}
\end{equation}
where $\omega_i$ is the weight assigned to direction $i$ given by the quadrature, in this case \citet{Carlson1963}.
Radiative heating contributed by the transition was calculated from \( J \) in regions where the optical thickness is relatively small, and from the divergence of the radiative energy flux (\( \nabla \cdot F \)) in optically thick regions, similar to the approach adopted for multigroup RT \citep{Bruls1999A&A}. The heating rate is given by
\begin{equation}
Q = h \nu_0 \left( n_l B_{lu} J - n_u A_{ul} \right) \mathrm{e}^{-\Delta \tau / \tau_0} - \nabla \cdot F \left( 1 - \mathrm{e}^{-\Delta \tau / \tau_0} \right), 
\label{eqn:Q-implement}
\end{equation}
where \( \tau_0 = 0.1 \), \( \Delta \tau = \chi \min(\Delta x, \Delta y, \Delta z) \), \( \chi \) is the opacity, and \( \Delta x \), \( \Delta y \), and \( \Delta z \) are the grid cell sizes in the three spatial directions. In optically thick regions, the terms \( n_l B_{lu} J \) and \( n_u A_{ul} \) are nearly equal, making round-off errors significant. In such cases, computing the heating rate from the radiative energy flux yields more accurate results.

\section{Results} \label{sec:result}

\begin{figure*}[ht]
    \centering
    \includegraphics[width=1.\linewidth]{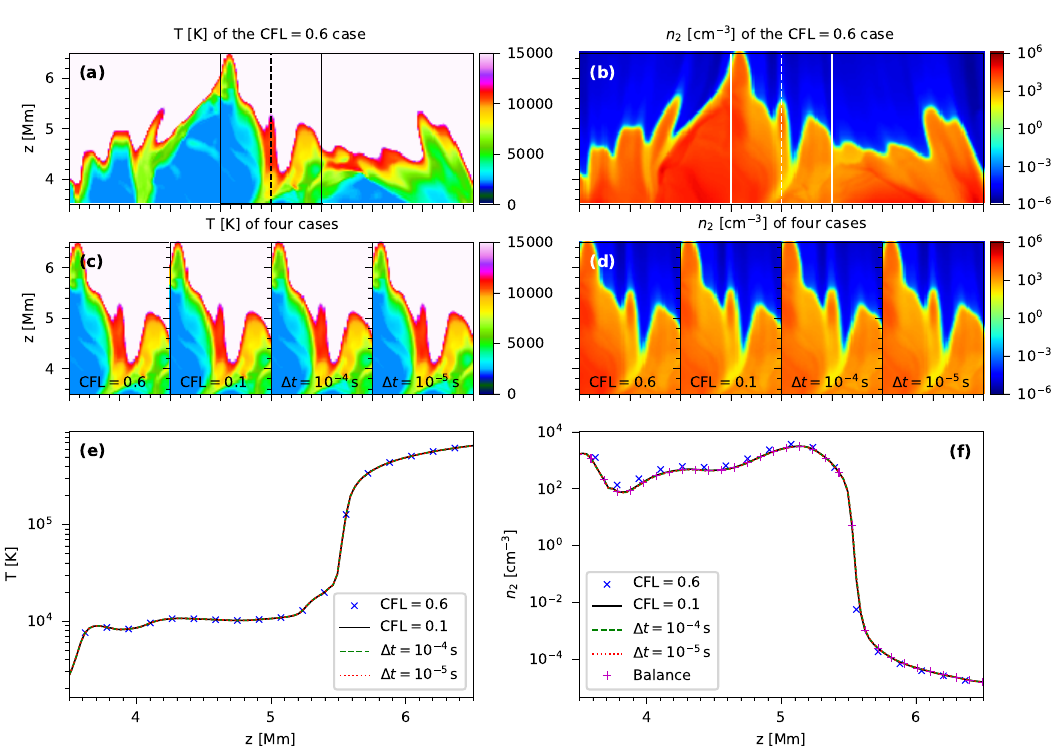}
    \caption{Temperature and \(n_2\) population distributions at \(t = 60\,\mathrm{s}\) for the four simulation cases. The average time steps of the CFL = 0.6 and CFL = 0.1 cases are approximately $10^{-2}\,\mathrm{s}$ and $2 \times 10^{-3}\,\mathrm{s}$, respectively. Panels (a) and (b) display the temperature and \(n_2\) distributions for the case with CFL = 0.6. Panels (c) and (d) show the corresponding regions (outlined in panels (a) and (b)) for all four cases, respectively. Panels (e) and (f) present 1D profiles of temperature and \(n_2\), respectively, extracted along the dashed lines in panels (a) and (b), for all four cases. In panel (f), an additional \(n_2\) profile computed from the balance solution for the \(\Delta t = 10^{-5}\,\mathrm{s}\) case is also shown for reference.}
    \label{fig:timestep}
\end{figure*}

\subsection{The influence of time step selection on the accuracy of simulation results} \label{sec:dt_survey}

In this subsection, we investigate the influence of time step selection on the evolution of the level populations in simulations.
Radiative excitation and de-excitation are processes that occur on very short timescales. For example, the radiative de-excitation timescale for the Lyman-$\alpha$ transition is given by $1/R_{21} \approx 1/A_{21} \approx 2\,\mathrm{ns}$. Resolving the radiative de-excitation process in simulations would require a time step of $\Delta t < 1\,\mathrm{ns}$, which is computationally prohibitive for a 3D MHD simulation, where timesteps are typically on the order of $10^{-3}$ to $10^{-1}$ seconds.
However, due to this extremely short timescale, the radiative relaxation of the level populations occurs rapidly, such that the population distribution can be considered to be in balance, dependent on the local radiation field, at each moment.
To capture the evolution of the level populations, a time step on the timescale of radiation field variations is required, rather than one based on the nanosecond timescales of radiative excitation and de-excitation. 
The timescale of radiation field variations should be comparable to that of MHD processes, such as convection, as population evolution dominated by collisions occurs over longer timescales. According to \citet{Carlsson2002ApJ}, the ionization/relaxation timescale associated with chromospheric shocks range from 10 to $10^3$ seconds, increasing dramatically in the 
lower, cooler layers of the chromosphere ($10^3$ to $10^5$ seconds).
Consequently, the evolution of the level populations can, in principle, be captured using the MHD time step in the simulation. To verify this, we evaluated the convergence of the simulation results by varying the time step size.

The survey comprises four 2D simulation cases, all of which begin with the same initial condition. In two of these cases, the time step is determined by the Courant–Friedrichs–Lewy (CFL) condition, with CFL numbers set to 0.6 and 0.1, respectively. The resulting average time steps are approximately $10^{-2}\,\mathrm{s}$ and $2 \times 10^{-3}\,\mathrm{s}$. In the remaining two cases, the time step is explicitly specified as $10^{-4}\,\mathrm{s}$ and $10^{-5}\,\mathrm{s}$, respectively. 
The initial condition originates from the output generated by the MURaM chromospheric extension. The simulation domain spans $\rm 8.192\ Mm \times 8.192\ Mm$, which is covered by $256 \times 256$ grid points. This corresponds to a resolution of 32 km both horizontally and vertically.  The lower boundary is located in the convection zone, the upper boundary is in the corona.
A Lyman-$\alpha$ bandwidth of $W_{\nu}=1\times 10^{12}\ \mathrm{Hz}$ and a Lyman-$\beta$ bandwidth of $W_{\nu}=1.25\times 10^{12}\ \mathrm{Hz}$ are adopted, as described in Sec.~\ref{sec:model-RT}. Radiative transfer is solved for both the corona and the chromosphere, with the parameter $\Delta \tau_{\mathrm{min}}=10^{-8}$ (see Sect.~\ref{sec:numerical-RT}) used in MURaM.

The temperature and $n_2$ population number density distributions at $t = 60\,\mathrm{s}$ are compared in Fig.~\ref{fig:timestep}. 
The four cases exhibit similar temperature and population distributions, as is shown in the figure. The profiles in Fig.~\ref{fig:timestep} (e)-(f) show a slight change in the results when the CFL number is reduced from 0.6 to 0.1, with no noticeable effect from further reductions in time step size. As expected, the population evolution is well captured when using the MHD time step. 
A balance solution for \( n_2 \) in the \( \Delta t = 10^{-5} \) case is shown in Fig.~\ref{fig:timestep} (f), which is given by 
\begin{equation}
    n_2 = \frac{\sum_{j \neq 2} n_{j} (C_{j2} + R_{j2})}{\sum_{j \neq 2} (C_{2j}+R_{2j})} . \label{eqn:equilibrium}
\end{equation}
The simulated \( n_2 \) distribution is in good agreement with the balance solution, as expected.

The temporal evolution of level populations can be classified into fast and slow evolution \citep{Carlsson2002ApJ,Judge2005JQSRT}. The slow evolution corresponds to changes occurring under approximate balance, a condition typical in the chromosphere. In contrast, the fast evolution describes the rapid establishment of approximate balance from an out-of-balance state, driven by the radiative transitions. 
In the code, the time evolution is solved using the Newton--Raphson method, which captures the faster rates, and allows for accurate modeling of the slow rates, playing a crucial role in the accurate modeling of population dynamics \citep{Przybylski2022A&A}. 
The iterative form of the rate equation Eqn.~(\ref{eqn:Rate}) is given by
\begin{equation}
f_i = \frac{n_i^{t_0+\Delta t}}{n_i^{t_0}} - \frac{\Delta t}{n_i^{t_0}} \left[ \sum_{j \neq i} n_{j} (C_{ji} + R_{ji}) - n_{i} \sum_{j \neq i} (C_{ij}+R_{ij}) \right] - 1 =0,
\end{equation}
where the advection term is neglected.
In the case of slow evolution -- that is, when the timescale of change is much larger than the time step, $\Delta t$ -- the behavior of this method is consistent with that of the explicit method.
For fast evolution, where the timescale is much shorter than $\Delta t$, the populations will reach an approximate balance given by
\begin{equation}
\sum_{j \neq i} n_{j} (C_{ji} + R_{ji}) - n_{i} \sum_{j \neq i} (C_{ij}+R_{ij}) \approx 0,
\end{equation}
within one time step, in agreement with the actual physical process.
For more detailed study and discussion of population evolution, refer to \citet{Carlsson2002ApJ} and \citet{Judge2005JQSRT}.

\subsection{Accuracy of the single-frequency approach for the Lyman-$\alpha$ and $\beta$ transitions} \label{sec:one-frequency}

To assess the accuracy of the single-frequency approach applied to the Lyman lines, we compare its results with reference solutions obtained using a more detailed multifrequency approach.
Using the output of the CFL = 0.6 case presented in Sect.~\ref{sec:dt_survey}, we computed the radiative heating and cooling rates, as well as the radiative rate coefficients, with both methods.

\begin{figure*}[htp]
    \centering
    \includegraphics[width=0.99\linewidth]{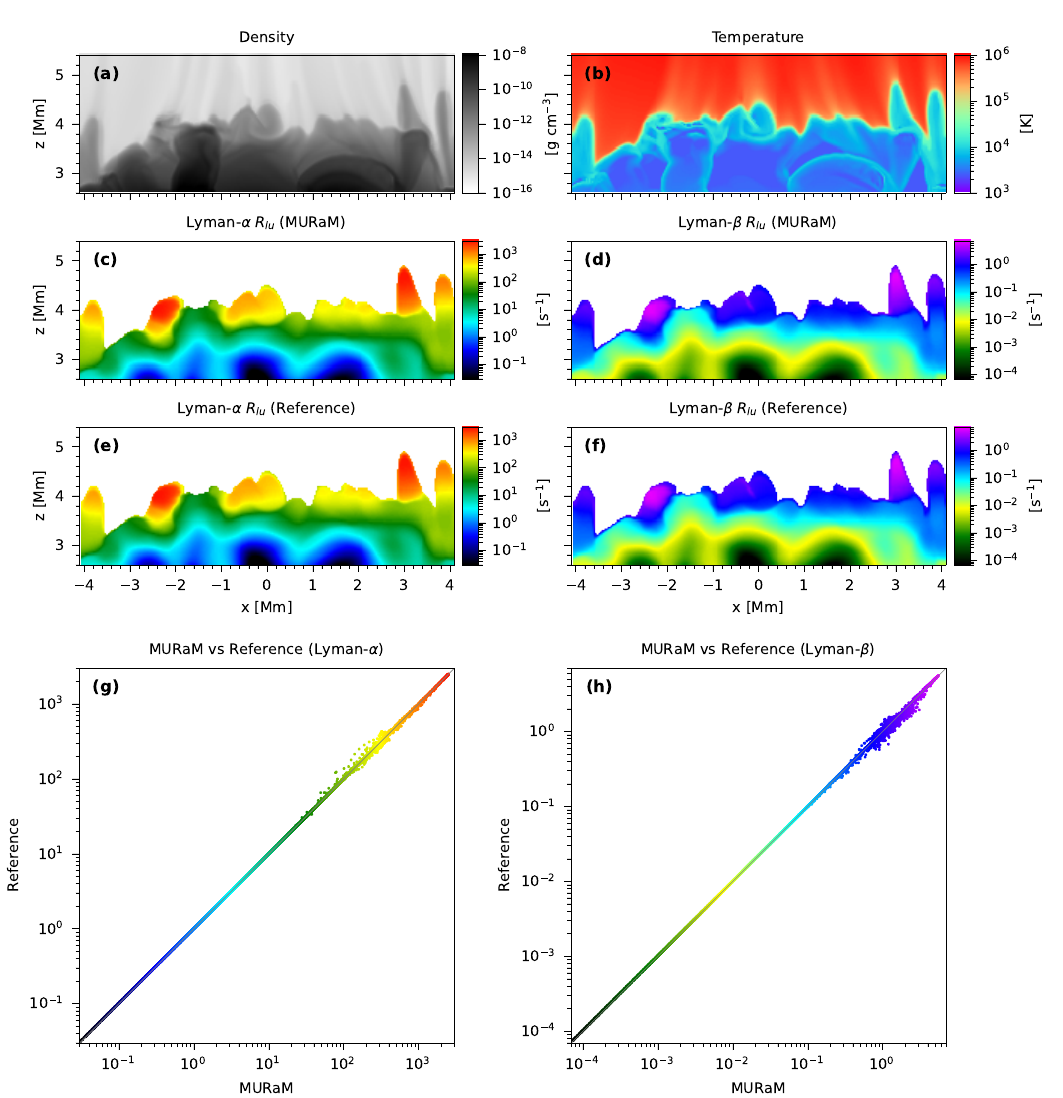}
    \caption{Comparison of the upward radiative rate coefficients obtained by MURaM (using the single-frequency approach) and reference (using the multifrequency approach by Lightweaver). Panels (a) and (b) show the atmospheric density and temperature distributions, respectively. 
    Panels (c) and (d) display the Lyman-$\alpha$ and Lyman-$\beta$ rate coefficients in the cool region ($T<=0.2\ \mathrm{MK}$) computed by MURaM, while panels (e) and (f) present the corresponding coefficients from the reference. 
    Panels (g) and (h) display the results of the point-to-point amplitude comparison, with the horizontal axis representing the values from MURaM and the vertical axis corresponding to the amplitudes from the reference. The points are colored according to their corresponding color in the MURaM result panels (c, d), so the area where the points are located can be inferred from the color.
    Since the highly ionized corona is not the region of interest for NLTE RT, the rate coefficients at high temperatures ($T>0.2\ \mathrm{MK}$) were manually set to zero to emphasize the comparison results in the lower atmosphere.
    The time corresponding to the simulated data is $t = 1\ \mathrm{s}$.}
    \label{fig:Rlu_MURaM+LW}
\end{figure*}

\begin{figure*}[htp]
    \centering
    \includegraphics[width=1.\linewidth]{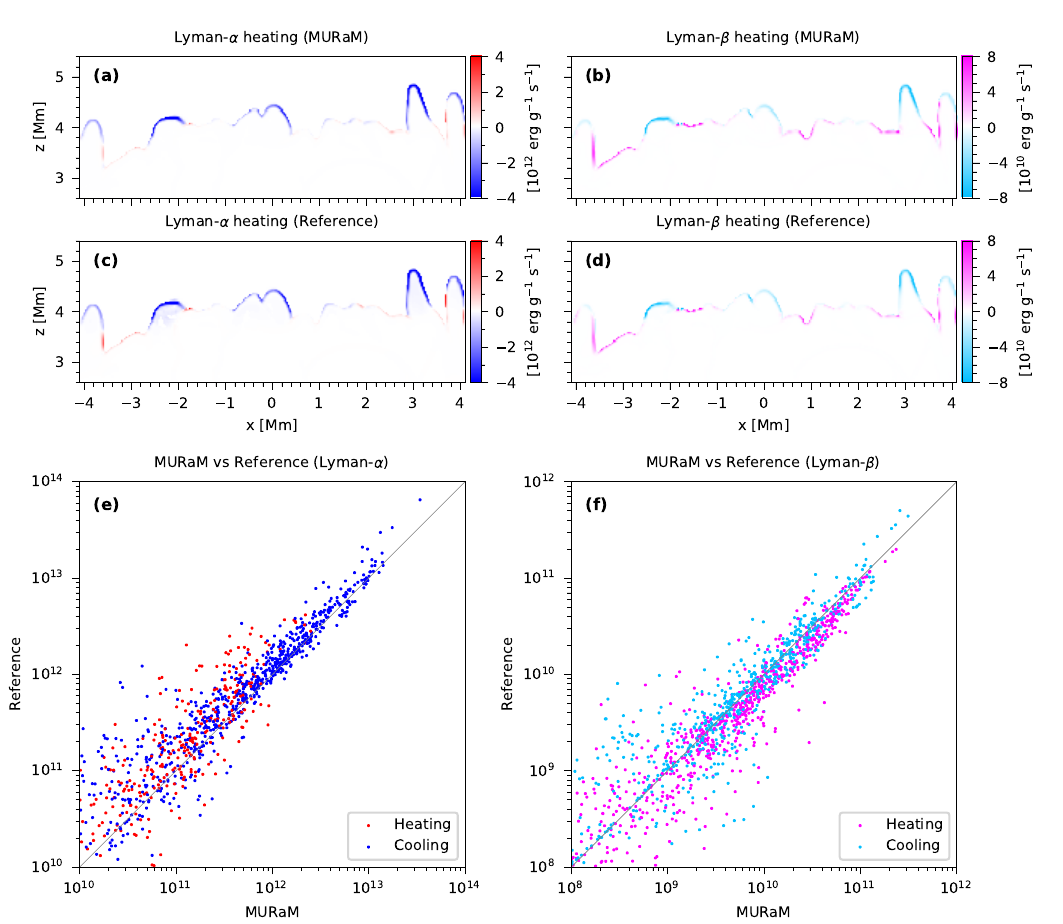}
    \caption{Comparison of the upward radiative rate coefficients obtained by MURaM and Lightweaver (reference). Panels (a) and (b) show the heating effects contributed by the Lyman-$\alpha$ and Lyman-$\beta$ lines from MURaM, while panels (c) and (d) present the corresponding heating effects in the reference. 
    Panels (e) and (f) give the point-to-point amplitude comparison, where the horizontal axis corresponds to the values obtained from MURaM and the vertical axis to the amplitudes from the reference. The points are colored based on their sign in both the MURaM and reference results, with red and pink indicating heating, and blue and cyan indicating cooling.
    The time corresponding to the simulated data is $t = 1\ \mathrm{s}$.}
    \label{fig:Q_MURaM+LW}
\end{figure*}

The single-frequency results were obtained from the MURaM code, while the multifrequency reference results were generated using the Lightweaver framework \citep{Osborne2021ApJ}.
In the single-frequency approach, only the hydrogen level population data were used in the computation with a specified line width.
In the multifrequency calculations, temperature data were additionally used to model spectral line broadening effects. 
By setting the velocity to zero, the Doppler effect is neglected. Sect.~\ref{sec:solver} demonstrates that it has an insignificant impact on the radiative rates. Therefore this simplification has a negligible impact on our results.
Lightweaver solves the RT equation using the short-characteristics approach for 1D and 2D problems. It accounts for partial frequency redistribution (PRD) in spectral lines and computes the specific intensity at each frequency, spatial point, and direction. 
The Lyman-$\alpha$ (line center at 121.5684 nm) multifrequency results are computed using 99 frequency points within a wavelength range of 120.8384 – 122.2983 nm, while the Lyman-$\beta$ (line center at 102.5733 nm) results are obtained from 48 frequency points within a wavelength range of 102.3167 – 102.8299 nm. 
For consistency, the RT equation was solved using the second-order solver BESSER, the same solver used in the MURaM code. A 24-bin spherical space discretization was employed in solving the equation, as in the MURaM code, but with a different quadrature provided by \citet{Stepan2020A&A}.
The multifrequency radiative heating or cooling rates and radiative rate coefficients were first computed on the RT mesh, then interpolated onto the MHD mesh, and subsequently compared to the single-frequency results at the MHD mesh, which were computed using the method described in Sec.~\ref{sec:numerical-RT}.

The results of the single-frequency approach and the multifrequency reference calculations are presented and compared in Fig.~\ref{fig:Rlu_MURaM+LW} and \ref{fig:Q_MURaM+LW}. 
As is shown in Fig.~\ref{fig:Rlu_MURaM+LW} (c)-(f), the spatial distribution of upward radiative rate coefficients ($R_{lu}$) of two Lyman lines are well reproduced by the single-frequency approach. The point-to-point comparison in Fig.~\ref{fig:Rlu_MURaM+LW} (g)–(h) shows that the rate coefficient amplitudes are in excellent agreement across most of the locations, with only minor deviations observed. These minor discrepancies are located at the upper chromosphere and the transition region, where the medium transitions from optically thick to optically thin. Most regions of the chromosphere are optically thick in  Lyman-$\alpha$, allowing the single-frequency approach to yield accurate upward rate coefficients in these areas. 
The amplitude and spatial distribution of the heating and cooling effects are also well captured, as is shown in Fig.~\ref{fig:Q_MURaM+LW}. In terms of correlation with the reference, the heating or cooling effect is not captured as accurately as the radiative rate coefficient. Throughout most of the chromosphere, the rates of photon emission and absorption are nearly balanced in the transitions, such that small deviations in the radiative rate coefficient can lead to proportionally larger variations in radiative heating or cooling.
The strongest radiative heating and cooling effects occur at the transition region boundary.
The detailed radiative balance of the lines breaks down there, refer to Fig.~\ref{fig:balance_old+new} and Sect.~\ref{sec:compare-versions}. 

The spatial distribution of the heating and cooling effects of the Lyman-$\beta$ line is similar to that of the Lyman-$\alpha$ line, but its amplitude is two orders of magnitude weaker. 
The downward radiative rate coefficients, $R_{ul}$, are very close to the constants $A_{ul}$ in the lower atmosphere, as the radiative de-excitation of the Lyman lines in this region primarily governed by the spontaneous de-excitation process. Therefore, the results are not compared here.

\subsection{Effects of the NLTE radiative transfer treatment of the Lyman series} \label{sec:compare-versions}

We now evaluate the impact of employing NLTE RT in the treatment of the Lyman series on chromospheric simulations. We conducted simulations using both the updated version of the MURaM code and the previous version, as it is described in \citet{Przybylski2022A&A}, and compared the results. As is described in Sect.~\ref{sec:method}, the previous version assumes detailed radiative balance for the Lyman series by setting the radiative rates to zero in the implementation, and computes radiative cooling from hydrogen using the recipe from \citet{Carlsson2012A&A}. In contrast, the updated version solves the NLTE RT equations to calculate the radiative rate coefficients for the Lyman-$\alpha$ and Lyman-$\beta$ transitions, as well as the associated radiative heating and cooling rates. More details are provided in Sect.~\ref{sec:pops-equations}.
Both simulations use the same initial condition, as used in Sect.~\ref{sec:dt_survey}.
A CFL number of 0.6 is used in both simulations to limit the timestep.

\begin{figure*}[htp]
    \centering
    \includegraphics[width=1.\linewidth]{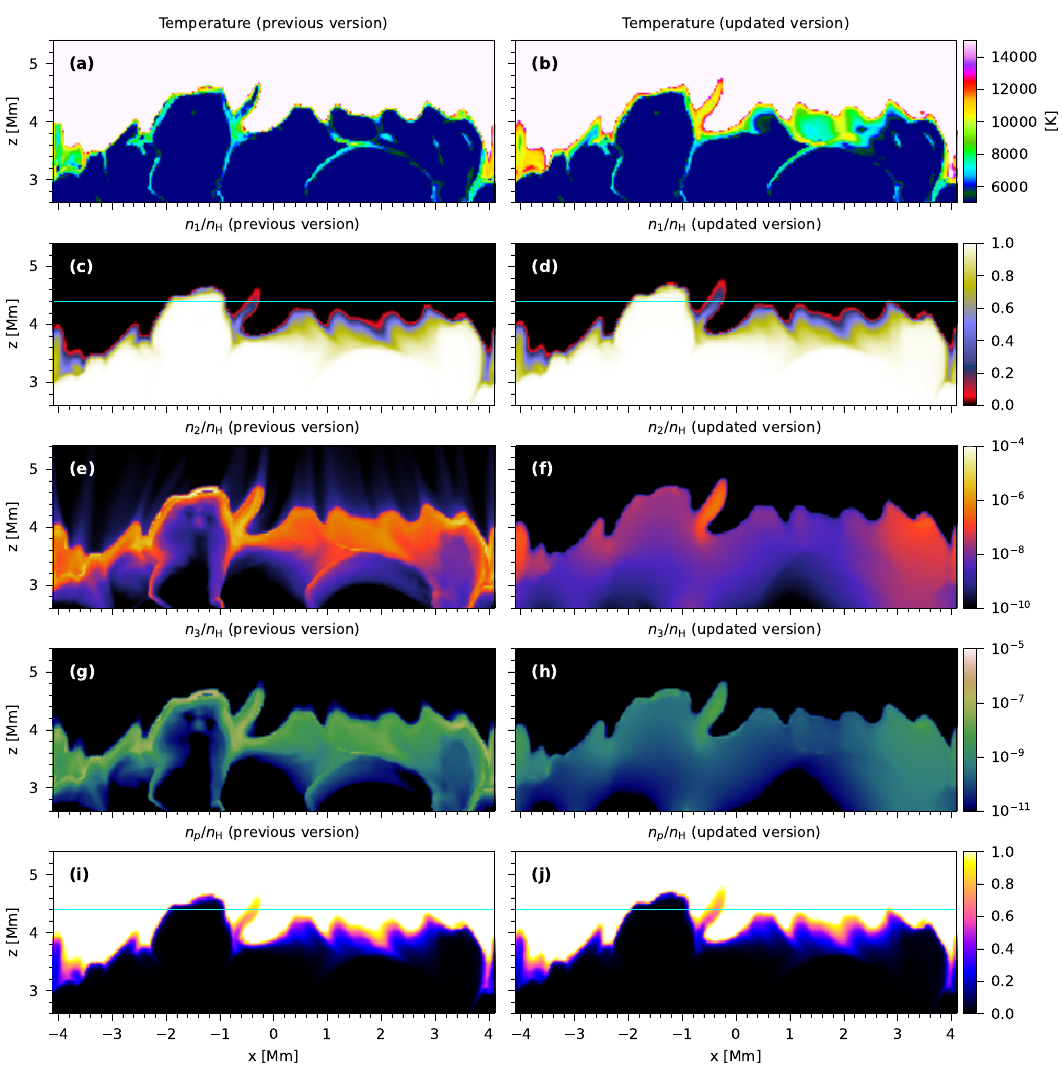}
    \caption{Comparison of the temperature and population distributions obtained by two versions of the MURaM code. 
    The left column shows results from the previous version, while the right column presents results from the updated version. 
    The first row displays the temperature distribution, while the second, third, fourth, and fifth rows show the population fractions of level~1, level~2, level~3, and protons, respectively, where $n_\mathrm{p}$ is the proton number density and $n_\mathrm{H}$ is the total hydrogen number density.
    To compare the extended height of the incompletely ionized region, a reference line is provided in panels (c), (d), (i) and (j).
    The previous version assumes $R_{lu}=R_{ul}=0$ for the Lyman series and computes radiative cooling using an empirical recipe, whereas the updated version calculates the Lyman-$\alpha$ and $\beta$ lines rate coefficients and the associated radiative heating and cooling through solving the RT equation.
    The time corresponding to the simulated data is $t = 20\, \mathrm{s}$}
    \label{fig:pops_old+new}
\end{figure*}

The chromosphere produced by the two versions of the code differ significantly, particularly in terms of temperature and hydrogen level populations. 
Fig.~\ref{fig:pops_old+new} presents a comparison of the temperature distributions and hydrogen level population fraction distributions from the two simulations at $t=20\ \mathrm{s}$.
In terms of temperature, the updated version of the code produces a hotter upper chromosphere. Given that the two versions employ different methods to compute radiative cooling, differences in temperature are expected. 
Additionally, differences in the population states will change the amount of internal energy held in ionization and excitation, changing the temperature.
From a physical perspective, the previous version accounts only for radiative cooling and neglects radiative heating related to the absorption of Lyman-$\alpha$ at the wavefront \citep{Carlsson2012A&A}. Therefore, it is reasonable that the updated version, which includes both processes, yields a higher chromospheric temperature. The radiative cooling due to the Lyman continuum and H-$\alpha$ have not been included in the update version, which may also contribute to the observed temperature differences. 

\begin{figure*}[htp]
    \centering
    \includegraphics[width=1.\linewidth]{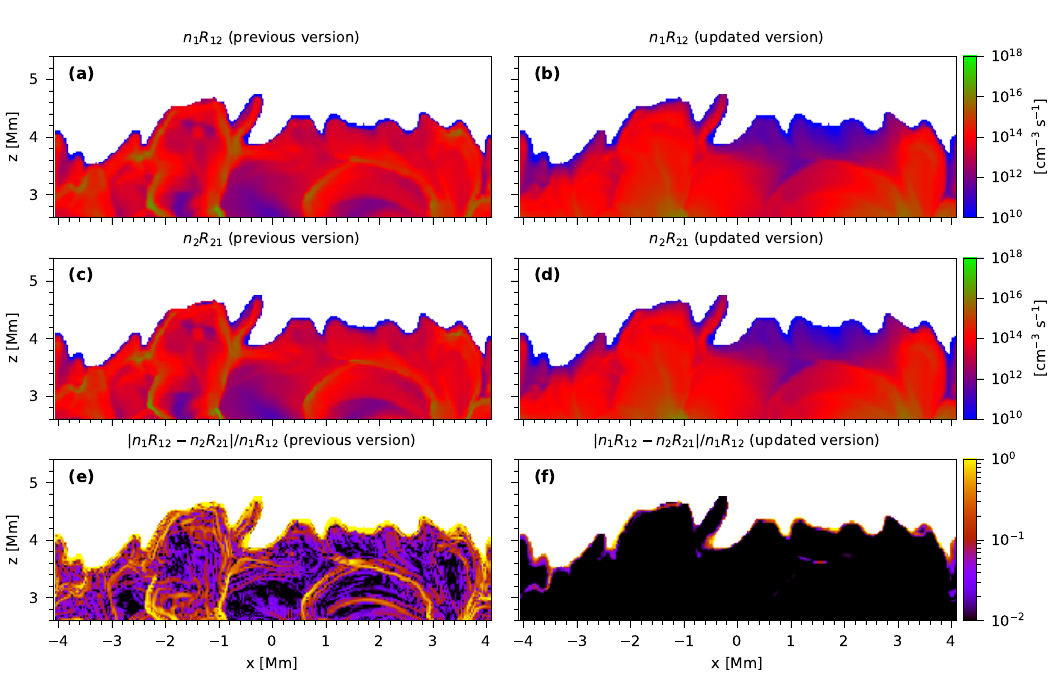}
    \caption{Comparison of the degree to which the population distributions provided by the two versions of the MURaM code conform to the radiative balance. Radiative upward and downward transition rates are computed from the populations shown in Fig.~\ref{fig:pops_old+new} using the Lightweaver code and are illustrated in panels (a)-(d).
    Panels (e) and (f) show the degree of deviation from the detailed radiative balance condition in the results from the two versions, with the calculation formula provided in the title.
    The time corresponding to the simulated data is $t = 20\, \mathrm{s}$.
    The rate coefficients at high temperatures ($T>0.2\ \mathrm{MK}$) were manually set to zero to emphasize the results in the lower atmosphere.}
    \label{fig:balance_old+new}
\end{figure*}

In terms of populations, noticeable differences between the two versions are observed. 
The ground level population fraction and ionization fraction distributions (Fig.~\ref{fig:pops_old+new} c, d, i, j) show that the incompletely ionized region reaches higher altitudes in the version incorporating Lyman-$\alpha$ radiation transfer (updated version), which aligns with the conclusions of \citet{Golding2016ApJ}. 
The spatial distributions of the populations of the upper level of the transitions, $n_2$ and $n_3$, are further decoupled from the temperature and exhibit smoother profiles in the updated version.
In the previous version, collision processes dominate excitation and de-excitation, since the radiative rate coefficients for the lines are set to zero. Since the collisional rate coefficients are temperature-dependent, some structures in the temperature distribution also appear in the distributions of $n_2$ and $n_3$. 
In the updated version, the inclusion of radiative de-excitation processes suppresses the occurrence of high $n_2$ and $n_3$ fractions in the chromosphere, resulting in reduced peak values. 
The inclusion of RT in the updated version introduces a new channel for energy exchange between different regions, potentially contributing to the differences in level populations.
The populations obtained in the updated version are closer to satisfying the detailed radiative balance condition, as demonstrated in Fig.~\ref{fig:balance_old+new}. In the results from the updated version, only the region near the wavefront deviates from detailed radiative balance, while the rest of the chromosphere can be approximated as being in detailed radiative balance. Despite adopting the assumption of detailed radiative balance, the previous version produces simulation outcomes that deviate markedly from detailed radiative balance condition.

Table~\ref{tab:speed} shows the impact of activating this NLTE RT module on simulation speed. The computational time increases by approximately 9\% to 21\%, depending on the number of spectral lines included and whether RT is resolved in the optically thin corona. Compared with Case 1, the increased computation time in the other cases is primarily attributable to the solution of the Lyman RT problem. There might remain potential to improve the computational efficiency of this module.

\begin{table*}[h]
\caption{Comparison of computational time consuming under different NLTE settings.}
\label{tab:speed}
\centering
\begin{tabular}{l c c c c}
\hline\hline
Case & Lyman-$\alpha$ & Lyman-$\beta$ & $\Delta\tau_{\textrm{min}}$ & Cost [$\mu$s per grid point per core per time step]  \\
\hline
1 & off & off & - & 38.70 \\
2 & on & off & 10$^{-5}$ & 42.14 \\
3 & on & on & 10$^{-5}$ & 45.35 \\
4 & on & off & 10$^{-8}$ & 42.72 \\
5 & on & on & 10$^{-8}$ & 46.88 \\
\hline
\end{tabular}
\tablefoot{Five simulation cases, all starting from the same initial conditions described in Sect.~\ref{sec:dt_survey}, are compared.
In Case 1, the NLTE RT module was not activated, while it was enabled in Cases 2 to 5. Radiative transfer was solved for Lyman-$\alpha$ only in Cases 2 and 4, and for both Lyman-$\alpha$ and Lyman-$\beta$ in Cases 3 and 5. In Cases 2 and 3, we restricted RT to the lower atmosphere by adopting a large value of $\Delta\tau_{\mathrm{min}}$, whereas in Cases 4 and 5 we solved it for both the lower atmosphere and the optically thin corona using a smaller $\Delta\tau_{\mathrm{min}}$. Each simulation was run for 1000 time steps, and the computational time was measured over time steps 100 to 1000. The first 100 time steps were excluded to allow for atmospheric relaxation.}
\end{table*}

\subsection{Impacts of the solver and the Doppler effect} \label{sec:solver}

\begin{figure*}[htp]
    \centering
    \includegraphics[width=1.\linewidth]{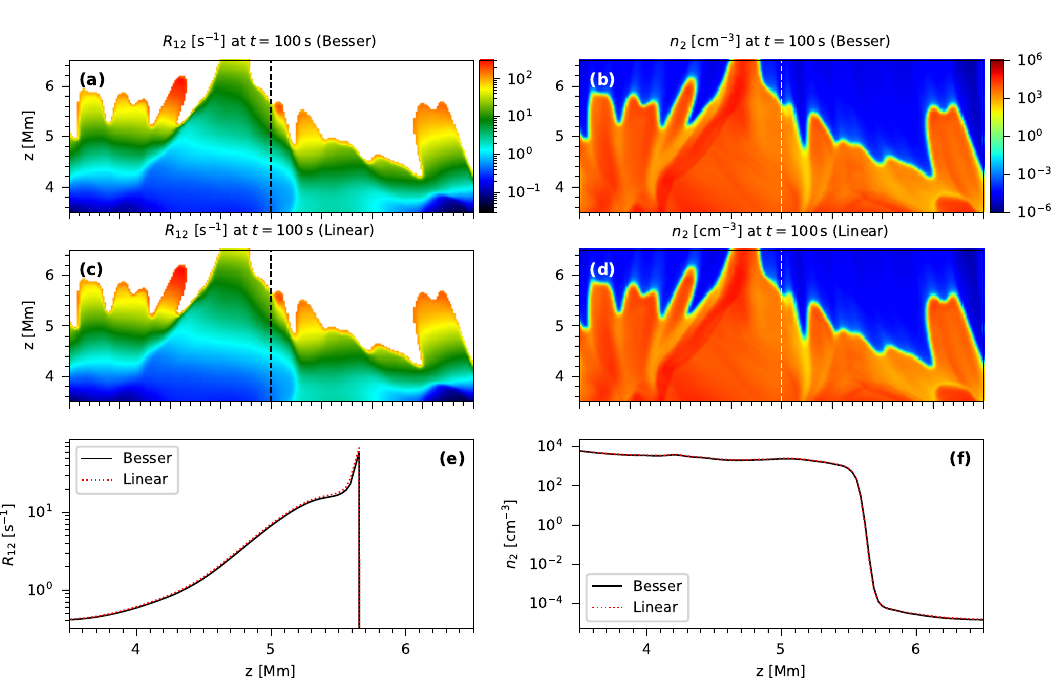}
    \caption{Comparison of the simulation results generated by two solvers. The radiative rate-coefficient distributions ($R_{12}$) produced by the BESSER and linear solvers are shown in panels (a) and (c), respectively, while panel (e) compares the distributions extracted along the dashed line depicted in panels (a) and (c). Similarly, the level-population distributions ($n_{2}$) obtained from the two solvers are presented in panels (b) and (d), respectively, with panel (f) showing the corresponding comparison along the dashed line. All results correspond to $t = 100$ s. The rate coefficients at high temperatures ($T>0.2\ \mathrm{MK}$) were manually set to zero to emphasize the results in the lower atmosphere.}
    \label{fig:solver}
\end{figure*}

Two schemes, the BESSER and the linear solver, were implemented for the module, as is detailed in Section \ref{sec:numerical-RT}. To evaluate their differences, we conducted a test comparing the simulation results produced by the two solvers. In the test, simulations were performed using the BESSER and linear solvers, respectively. Both the Lyman-$\alpha$ and Lyman-$\beta$ lines were activated in the simulations, with $\Delta\tau_{\mathrm{min}}$ set to 10$^{-8}$. Fig.~\ref{fig:solver} presents the radiative rate coefficient ($R_{12}$) and population distribution ($n_2$) at $t = 100$ s for both simulations. As is shown in panel (e), slight differences can be observed in the radiative rates obtained by the two solvers. However, this difference in radiative rate does not lead to significant long-term differences in the population distribution, as indicated by the nearly identical population profiles in panel (f). There may be an underlying mechanism that drives the convergence of the population evolution.

\begin{figure*}[htp]
    \centering
    \includegraphics[width=1.\linewidth]{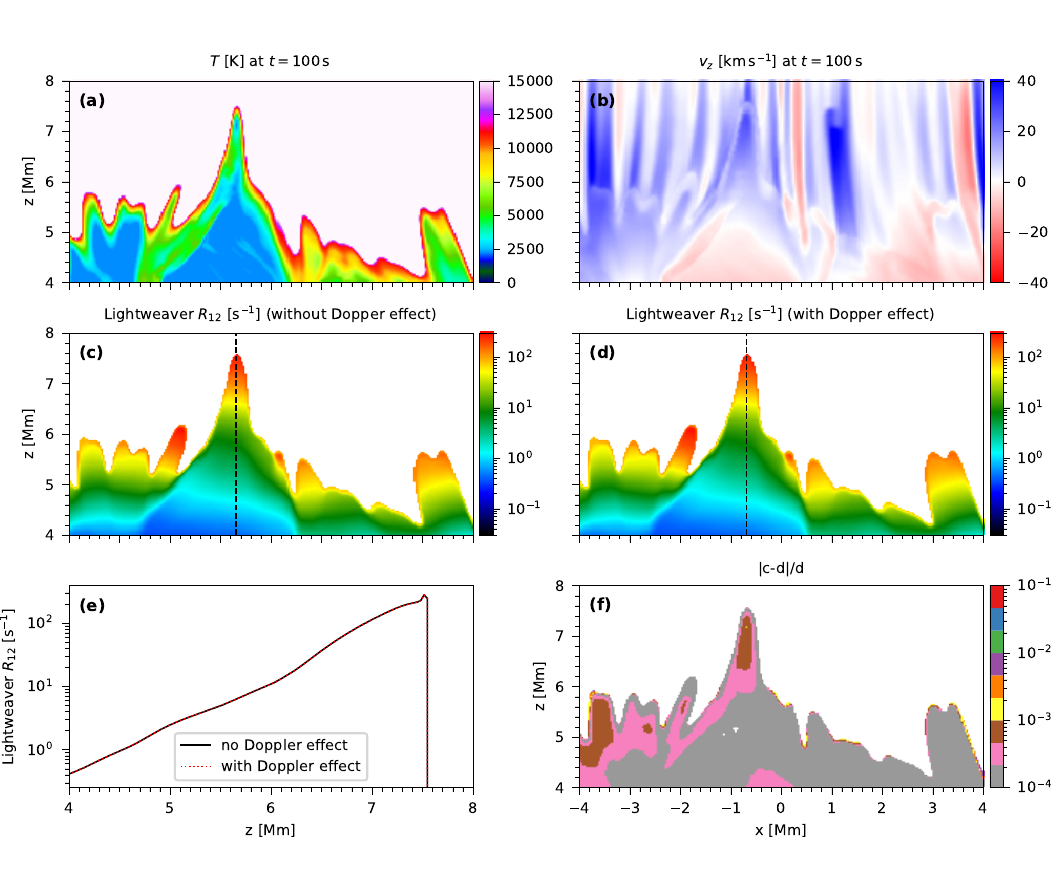}
    \caption{Comparison of the Lyman-$\alpha$ upward radiative rate coefficient ($R_{12}$) computed with and without the Doppler effect. Panels (a) and (b) present the simulation outputs for temperature and vertical velocity, respectively. Panels (c) and (d) show the rate coefficients calculated with and without accounting for the Doppler effect. Panel (e) compares the rate coefficients along the dashed line indicated in panels (c) and (d), while panel (f) presents the point-to-point comparison, with the evaluation formula provided above the panel. The rate coefficients at high temperatures ($T>0.2\ \mathrm{MK}$) were manually set to zero to emphasize the results in the lower atmosphere.}
    \label{fig:doppler}
\end{figure*}

In the single-frequency treatment of the Lyman lines, the Doppler effect is neglected. To assess its impact on the Lyman-$\alpha$ line, we conducted a test using Lightweaver to calculate the upward radiative rate coefficient ($R_{12}$) both with and without the Doppler effect, and subsequently compared the results. The calculations are based on the data produced by the BESSER solver, as described earlier in this chapter, using the hydrogen level populations, temperature and velocity values in the computation. For the calculations that neglect the Doppler effect, the plasma velocity was set to zero, whereas the calculations accounting for the Doppler effect employed the velocity values obtained from the simulation output. Fig.~\ref{fig:doppler} presents the simulation results, including temperature, vertical velocity, and the computed rate coefficients from both approaches. The results show that the difference in rate coefficients between the two methods is on the order of $10^{-3}$, indicating that the Doppler effect can be safely neglected in this context.

\section{Discussion and conclusion} \label{sec:summary}

To more accurately simulate chromospheric dynamics, we incorporated an NLTE RT module into the MURaM code, and used it to compute radiative rate coefficients and radiative heating or cooling associated with the Lyman-$\alpha$ and $\beta$ lines. In this module, the time-dependent RT equation is solved based on the evolving population densities within the simulation.
To reduce computational cost, a single-frequency approach was employed to treat these spectral lines, assuming uniform light intensity and opacity within a specified bandwidth for each line. To evaluate the accuracy of the single-frequency approach, the radiative rate coefficients and associated heating and cooling effects computed by MURaM were compared with reference solutions provided by the Lightweaver framework. The comparison results demonstrate that the radiative rate coefficients are accurately reproduced, and the radiative heating and cooling effects are also well captured.
After incorporating this module, the chromospheric simulation results from MURaM show significant changes, including a hotter upper chromosphere, an upward extension of incomplete ionization region, and hydrogen level populations more consistent with detailed radiative balance across the chromosphere.
The module enables simulations to accurately capture population evolution under MHD time steps, demonstrating its practicality for high-resolution multidimensional applications.

In future work, 3D simulations will be employed to further validate the module. Simulations will be performed using both the previous and updated versions, and the resulting atmospheric outputs (e.g., temperature, density, and velocity), as well as the structure of the jets, will be compared to evaluate the impact of the new module. The inclusion of RT for the Lyman-$\alpha$ and $\beta$ lines aims to enhance the accuracy of hydrogen population modeling in the chromosphere. 
To assess whether this objective has been successfully achieved, the forward-modeled results will be compared with imaging and spectroscopic observations in the Lyman-$\alpha$, Lyman-$\beta$, and H$\alpha$ bands. The MULTI3D tool will be used for imaging and spectroscopic synthesis \citep{Leenaarts2009ASPC}.
On the observational side, data from a variety of instruments can be used for comparison; for example, data from the Solar Ultraviolet Measurements of Emitted Radiation on board the Solar and Heliospheric Observatory (SOHO/SUMER, \citealp{Wilhelm1995SoPh}), the Extreme Ultraviolet Variability Experiment on board the Solar Dynamics Observatory (SDO/EVE, \citealp{Woods2012SoPh}), the Very high angular resolution ultraviolet telescope (VAULT2.0, \citealp{Vourlidas2016JAI}), the Spectral Imaging of the Coronal Environment on board the Solar Orbiter (SolO/SPICE, \citealp{SPICE2020A&A}), the Lyman-alpha Solar Telescope on board the Advanced Space-based Solar Observatory (ASO-S/LST, \citealp{Li2019RAA}), and the H$\alpha$ Imaging Spectrograph on board the Chinese H$\alpha$ Solar Explorer (CHASE/HIS, \citealp{Liu2022SCPMA}).\

In the single-frequency approach to RT, the bandwidth parameter ($W_{\nu}$) is a key factor. The default bandwidths for the Lyman-$\alpha$ and Lyman-$\beta$ lines were selected based on a limited parameter survey, based on how well the resulting upward radiative rate coefficients and heating or cooling effect matched those from Lightweaver. The test case is described in Sect.~\ref{sec:one-frequency}.
Our choice of Lyman-$\alpha$ line bandwidth differs from that of \citet{Golding2016ApJ}, who employed a bandwidth of $W_{\nu}=5\times 10^{11}\ \mathrm{Hz}$. This value is interpreted as the frequency range corresponding to half the maximum value of the Doppler-broadened line profile at a temperature of 10,000 K.
In our test, a Lyman-$\alpha$ bandwidth of around $W_{\nu}=1\times 10^{12}\ \mathrm{Hz}$ yields a good agreement with reference results. This value may physically correspond to the thermal broadening at a specific location within the wavefront or to the average thermal broadening across the wavefront region.
Bandwidth selection remains an area for improvement in our work. With a more extensive parameter survey, it may be possible to identify bandwidths that simultaneously capture accurate radiative rate coefficients and radiative heating and cooling effects in the upper chromosphere. The influence of simulation dimensionality and resolution on the optimal bandwidth also requires further investigation.

With regard to radiative cooling from hydrogen, only the contributions from the Lyman-$\alpha$ and $\beta$ lines are considered in the updated version of the MURaM code. According to \citet{Carlsson2012A&A}, these two spectral lines dominate the cooling across most of the chromospheric temperature range, with Lyman-$\alpha$ being the primary contributor. However, below temperatures of 7,000 K, H$\alpha$ becomes the dominant contributor to radiative cooling. In addition, the Lyman continuum also contributes significantly to chromospheric radiative losses. The impact of the absence of the H$\alpha$ and Lyman continuum on the evolution of chromospheric temperature remains to be investigated. 
Including RT for these lines is a potential direction for the future development of the MURaM code.
The current scheme could easily be extended to helium \citep{Golding2016ApJ}, as well as to \ion{Mg}{ii} and \ion{Ca}{ii}, to further improve the models.

\begin{acknowledgements}
      We would like to thank Narayanamurthy Smitha for helpful discussions related to RT and the solar chromosphere. This project has received funding from the European Research Council (ERC) under the European Union's Horizon 2020 research and innovation programme (grant agreement No. 101097844 — project WINSUN). The work of DP was funded by the Federal Ministry for Economic Affairs and Climate Action (BMWK) through the German Space Agency at DLR based on a decision of the German Bundestag (Funding code: 50OU2201). We gratefully acknowledge the computational resources provided by the Raven and Viper supercomputer systems of the Max Planck Computing and Data Facility (MPCDF) in Garching, Germany. We thank C. Osborne (Lightweaver) and A. Irwin (FreeEoS).
\end{acknowledgements}

\end{document}